\documentclass[10pt,conference]{IEEEtran} 
\IEEEoverridecommandlockouts
\usepackage{cite}
\usepackage{amsmath,amssymb,amsfonts}
\usepackage{algorithmic}
\usepackage{graphicx}
\usepackage{textcomp}
\usepackage{xcolor}
\def\BibTeX{{\rm B\kern-.05em{\sc i\kern-.025em b}\kern-.08em
    T\kern-.1667em\lower.7ex\hbox{E}\kern-.125emX}}
\usepackage[T1]{fontenc}
\usepackage{xcolor}
\usepackage{braket}
\usepackage{ulem}

\usepackage{verbatim} 
\usepackage{multirow}
\usepackage{url}

\newcommand{\htwo}{H$_2$\,}

\begin{document}

\title{Comparing performance of variational quantum algorithm simulations on HPC systems}

\author{
\IEEEauthorblockN{Marco De Pascale\IEEEauthorrefmark{1}, 
Tobias Valentin Bauer\IEEEauthorrefmark{1}, 
Yaknan John Gambo\IEEEauthorrefmark{1}\IEEEauthorrefmark{3}, 
Mario Hern\'{a}ndez Vera\IEEEauthorrefmark{1}, 
Stefan Huber\IEEEauthorrefmark{1},\\ 
Burak Mete\IEEEauthorrefmark{1}\IEEEauthorrefmark{3}, 
Amit Jamadagni\IEEEauthorrefmark{1}\textsuperscript{\textsection}, 
Amine Bentellis\IEEEauthorrefmark{2}, Marita Oliv\IEEEauthorrefmark{2}, Luigi Iapichino\IEEEauthorrefmark{1}, 
Jeanette Miriam Lorenz\IEEEauthorrefmark{2}
}
\IEEEauthorblockA{\IEEEauthorrefmark{1}Leibniz Supercomputing Centre of the Bavarian Academy of Sciences and Humanities (LRZ), \\ Garching b. München, Germany \\
Email:\{marco.depascale, tobias.bauer, yaknan.gambo, mario.hernandezvera, stefan.huber, burak.mete, luigi.iapichino\}@lrz.de}
\IEEEauthorblockA{\IEEEauthorrefmark{2}Fraunhofer Institute for Cognitive Systems IKS, München, Germany\\
Email: \{amine.bentellis, marita.oliv, jeanette.miriam.lorenz\}@iks.fraunhofer.de}
\IEEEauthorblockA{\IEEEauthorrefmark{3}Technical University of Munich, School of CIT, Department of Computer Science, 
Garching b. München, Germany}

}

\maketitle
\begingroup\renewcommand\thefootnote{\textsection}
\footnotetext{Current address: Oak Ridge National Laboratory, USA}
\endgroup

\begin{abstract}
Variational quantum algorithms are of special importance in the research on quantum computing applications because of their applicability to current Noisy Intermediate-Scale Quantum (NISQ) devices. The main building blocks of these algorithms (among them, the definition of the Hamiltonian and of the ansatz, the optimizer) define a relatively large parameter space, making the comparison of results and performance between different approaches and software simulators cumbersome and prone to errors. In this paper, we employ a generic description of the problem, in terms of both Hamiltonian and ansatz, to port a problem definition consistently among different simulators. Three use cases of relevance for current quantum hardware (ground state calculation for \htwo molecule, MaxCut, Travelling Salesman Problem) have been run on a set of HPC systems and software simulators to study the dependence of performance on the runtime environment, the scalability of the simulation codes and the mutual agreement of the physical results, respectively. The results show that our toolchain can successfully translate a problem definition between different simulators. On the other hand, variational algorithms are limited in their scaling by the long runtimes with respect to their memory footprint, so they expose limited parallelism to computation. This shortcoming is partially mitigated by using techniques like job arrays. The potential of the parser tool for exploring HPC performance and comparisons of results of variational algorithm simulations is highlighted.
\end{abstract}


\section{Introduction}
\label{intro}

On the way to unfolding the potential of Quantum Computing (QC), a key (and often underrated) role is played by the research on quantum algorithms capable of running on current hardware. We are in the era of Noisy Intermediate-Scale Quantum (NISQ) devices, and algorithmic research focuses on providing circuits capable of running on such devices. In the exploration of the algorithms and of the use cases where they are employed, a valuable tool is the simulation of quantum circuits on classical computing resources (see \cite{hxg22,xbs23,yse23} for recent reviews). Both the memory footprint and the computational complexity of such simulations make them a challenge that needs  to be tackled by using High Performance Computing (HPC) resources (e.g., \cite{djw19,Jones2019,Guerreschi2020}).

The variational quantum algorithms (VQAs) are an especially important category \cite{Cerezo2021,Fedorov2022}. With their shallow circuits, they are suitable to be run on NISQ systems. Moreover, the interplay between the QC part of their workflow and the classical optimization component is among the drivers of tighter integration of QC within the HPC environment, both at the hardware level, as well as for the software stack and programming environment \cite{hml21,srk22,ets23,set23,Schulz2023}. Unsurprisingly, the most promising early applications of quantum algorithms are based on VQAs, with the Variational Quantum Eigensolver (VQE) and the Quantum Approximate Optimisation Algorithm (QAOA) with their variants being the principal representatives. Simulating VQAs on classical resources presents some additional challenges concerning the simulation of non-parametrized circuits. VQA consists of a number of building blocks (definition of the Hamiltonian, state preparation and ansatz, parameter optimization), with active research on suitable recipes, depending on the use case, for each of them (for examples, see \cite{wrk22,omm22}). These different choices define a large parameter space and make direct comparisons between simulations cumbersome. 

In this work, we are interested in a specific scenario. Given a suite of use cases (which will be described in detail below), a comparison between different simulation packages on several HPC systems will be performed. The diagnostics of the comparison concern the performance of the simulations in a broad sense, including both figures of merit of the problem solution and the efficient usage of HPC resources. To address the consistency issues given by the large parameter space of the algorithms to be compared across different simulators, a parser has been developed. This tool allows a seamless porting of the Hamiltonian and ansatz throughout the simulators. Motivated by previous research \cite{bba21,jlh24}, runs performed on a “bare-metal” installation of the tools have been compared with runs done within a container since the latter is considered a viable deployment strategy for software on HPC systems.

For the comparison, we focus on three use cases, one from computational chemistry (the \htwo molecule) and two from optimization (MaxCut and the Traveling Salesman Problem). They all represent promising application areas on NISQ devices. For the sake of focus, the following design choices have been made in this study: only state vector simulators are tested, and compilation and transpilation of the resulting quantum circuits are ignored.  Some of the directions not taken mentioned above are indeed interesting and will be the subject of future research.

The paper is organized as follows: in Section \ref{sec:algorithm-and-use-cases}, the algorithms and use cases chosen for this study are described, and in Section \ref{sec:parser}, the developed Hamiltonian and ansatz parsers are introduced. In Sections \ref{sec:systems} and \ref{sec:sim}, we describe the HPC systems and the setup of the tests, respectively, while in Section \ref{sec:software}, the used simulations packages are listed. The results are presented in Section \ref{sec:results} and discussed in Section \ref{discussion}, where we also draw our conclusions.

\section{Algorithms and use cases}
\label{sec:algorithm-and-use-cases}

For this study, we have chosen a small sample of use cases that can be addressed by means of VQAs. This is different from other works that give priority to simulating scalable algorithms like the Quantum Fourier Transform \cite{ssa16} or to the execution of single-qubit gates \cite{Guerreschi2020,adp24}. The present choice, in our opinion, brings the spotlight to concrete and interesting applications for the user community.
In this Section, three different use cases from the domains of quantum chemistry and combinatorial optimization using two prevalent variational algorithms are presented: the simulation of the \htwo molecule, MaxCut, and the Traveling Salesperson Problem. In Section \ref{sec:results}, we discuss that the different problems are used to study different features of the simulations, both from the viewpoint of computational efficiency and of the solution quality.

\subsection{Algorithms}
\label{sec:algorithms}

\subsubsection{Variational Quantum Eigensolver}
The Variational Quantum Eigensolver (VQE), first described by \cite{peruzzo2014variational}, 
is a heuristic quantum-classical algorithm to find the minimum of a cost function. This cost function $C(\theta)$ is implemented as the expectation value of an observable $O$ in a quantum state $\ket{\Psi(\theta)}$ that is prepared by a parameterized quantum circuit (PQC), also called ansatz:

\begin{equation}
    C(\theta) = \bra{\Psi(\theta)} O \ket{\Psi(\theta)}
\end{equation}

A classical optimization routine is used to adapt the parameters of the PQC to find the optimum of the cost function.

\subsubsection{Quantum Approximate Optimization Algorithm}
One part of our work benchmarked quantum simulators with the Quantum Approximate Optimization Algorithm (QAOA) \cite{farhiQuantumApproximateOptimization2014}. A formalization can be found in \cite{hadfieldQuantumApproximateOptimization2019}.
	
We provide a brief description of the algorithm as follows: let $j$ be the number of layers applied in the QAOA algorithm, and $M$ be the number of qubits.
	
\begin{enumerate}
	\item 	A qubit-register is initialized as $|0\rangle \in (\mathbb{C}^2)^{\otimes M}$. 
	\item 	For each layer $j$, a phase-separation unitary $U_P(\alpha_j)$ depending on the cost function and a parameter vector $\alpha$ is applied. 
  	\item 	For each layer $j$, a mixing unitary $U_M(\beta_j)$ depending on the domain of possible solutions and a parameter-vector $\beta_j$ is applied.

\end{enumerate}
	
Together with the application of a defined number of layers and measurement, classical minimizers are used to minimize the energy by variation of the parameters $\alpha_j, \beta_j$ of each layer. 

\subsection{Use cases}
\label{sec:use-cases}

\subsubsection{Molecule simulation - \htwo}
\label{sec:molecule-simulation}
A basic computational primitive in quantum chemistry is the calculation of the ground state energy of molecules. The hydrogen molecule, with a basic electronic Hamiltonian, is often used to prototype and test new algorithms like VQE. The expectation value of the molecular Hamiltonian defines the cost function of VQE for quantum chemistry applications. The molecular hydrogen Hamiltonian is calculated in our benchmark in the second quantization formulation within the Born-Oppenheimer approximation and using the STO-3G Gaussian basis set. Subsequently, a Jordan-Wigner transformation is applied to obtain a new representation of the Hamiltonian given in terms of Pauli operators acting on four qubits. The classical optimizer implemented by the Scipy Python optimization library will invoke the cost function with the default Broyden-Fletcher-Goldfarb-Shanno (BFGS) algorithm \cite{bfgs2006}. Additionally, the trial quantum states, representing the wave function of the molecule, are prepared in this work using the Unitary Coupled-Cluster Singles and Doubles (UCCSD) circuit ansatz \cite{BARTLETT1989133}, which is consistently applied to the initial Hartree-Fock state reference.

\vspace{1mm}
\textit{Optimization problems:}
\label{sec:optimization-problem}
Two primary candidates were selected for investigation in the realm of combinatorial optimization. The first one is the MaxCut problem, a mathematical problem that, while extensively studied, exhibits limited practical utility within real-world contexts. The second one is the Traveling Salesperson Problem (TSP), which is also a key graph problem that has plenty of applicability in domains like logistics or planning.

From an implementation standpoint, the two problems can be solved similarly. It begins by mathematically defining the problem through a Quadratic Unconstrained Binary Optimization (QUBO). Given a matrix $Q$ representing the problem, a QUBO problem is formulated as follows:

\[\min\{x^{T}Qx\} \]
\[x \in \{0,1\}^n\]

By mapping the classical binary variables to operators, we transform the classical QUBO to the
qubit space, further resulting in an Ising-like Hamiltonian. The groundstate of the above obtained 
Ising Hamiltonian represents the optimal solution, and therefore, we minimize the energy by employing
variational quantum algorithms with various ansatzes, in this case, QAOA for the MaxCut problem and VQE for the TSP. In the following, we briefly
describe the classical QUBO formulation and the corresponding Ising Hamiltonian for both 
the MaxCut and TSP problem instances.

\subsubsection{MaxCut}
\label{sec:maxcut}
In graph theory, a maximum cut (for short MaxCut) is a cut containing at least as many edges as any other cut. To define the problem formally, we consider an undirected graph $G = (V, E)$ and define the cut set as follows
\begin{equation}
    \delta(U) = \{(u,v) \in E : u \in U, v \notin U\}\space\text{ with }U \subseteq V.
\end{equation}
Thus, the maximum cut is given by $\max\{|\delta(U)|\}$. In other words, we can associate 
a binary variable to every edge, say $x_{uv}$, connecting the vertices, $u, v$ and map the binary 
variable to 1 if in $\delta(U)$ and else 0, resulting in the corresponding QUBO formalism given by
\begin{equation}
C = \sum_{e\in E}x_{e},
\end{equation}
where the aim is to find a cut that maximizes C. The corresponding Ising Hamiltonian can be obtained
by mapping the binary variables to the projectors $\frac{1 - \sigma_{z}^{u}\sigma_{z}^{v}}{2}$.

\subsubsection{TSP}
\label{sec:tsp}
Given a complete, undirected, and weighted graph, the objective of the TSP problem is to determine the
shortest cycle that visits all vertices exactly once. Its symmetric adjacency matrix \( D \), which describes the distances between all pairs of vertices, provides a concise representation of a TSP instance. The QUBO formalism involving the binary decision variables $x_{ij}$ where $i$ and $j$ are the vertices in the graph is given by 
\begin{align}
\begin{split}
    C(\{x_{ij}\}) = \left[ \sum_{i,i',j = 1}^n D_{ii'} x_{ij} x_{i'(j+1)}\right.\\
     \left.+ P \sum_{i=1}^n \left(1 - \sum_{j=1}^n x_{ij}\right)^2\right. \\
     \left. + P \sum_{j=1}^n \left(1 -  \sum_{i=1}^n x_{ij}\right)^2 \right].
     \end{split}
\end{align}
In the above, P is a penalty factor for the two constraints, i.e., each vertex is visited exactly
once, and each position is asserted exactly once. The corresponding QUBO matrix element $Q_{ijkl}$, the interaction between $x_{ij}$ and $x_{kl}$, is given by

\begin{align}\label{eq: QUBO tensor}
\begin{split}
    Q_{ijkl} = -2sP\delta_{ik}\delta_{jl} + sP [\delta_{ik}(1-\delta_{jl}) + \delta_{jl}(1-\delta_{ik})] \\
    + sD_{ik} [\delta_{j(l+1)} + \delta_{j(l-1)}].
\end{split}
\end{align}
The first term describes the diagonal elements, the second term represents the penalty for constraint violations, and the third term corresponds to the TSP path length. With the QUBO matrix described, we can translate the problem into an Ising Hamiltonian by setting the variables to projectors, as discussed earlier, resulting in
\begin{equation}
    H_P = \sum_{(i,j) \neq (k,l)} \frac{Q_{ijkl}}{4}\sigma_z^{ij} \sigma_z^{kl} + \sum_{i,j,k,l} \frac{Q_{ijkl}}{2} \sigma_z^{kl} .
\end{equation}

\section{The Hamiltonian and the ansatz parsers}
\label{sec:parser}

One important aspect of benchmarking simulators is the consistency across different components of the benchmarks. In quantum algorithms, there could be several factors that could disrupt the consistency. The differences in Hamiltonian and ansatz generation, which are strictly quantum components, could result in substantial contrasts among various simulators for a subset of algorithms. To elucidate, the Hamiltonian is the energy operator of a quantum system and is primarily used in quantum algorithms to estimate the system's energy. The smallest eigenvalue of this operator represents the \textit{ground state energy} of that particular system, and preparing that state is the end goal for algorithms such as VQE and QAOA. For instance, for an energy estimation of an electronic structure, the operator is a \textit{molecular Hamiltonian}, mapped from a \textit{fermionic Hamiltonian} using methods such as the Jordan-Wigner transformation \cite{jordan1993paulische}. However, different simulators have their own implementations of such algorithms, and even though they implement the exact methods, the final results could include minor differences. For instance, we have realized that the molecular Hamiltonians generated in PennyLane and Qiskit simulators differ in the inclusion of nuclei repulsion, which adds a constant shift to the respective energies. Moreover, simulators may represent a problem’s ansatz in various ways, using different gates or gate counts depending on their native gate sets or pre-compiled circuit definitions.

To circumvent this problem, we propose to use two simulator agnostic intermediate representations (IR), one for the Hamiltonian and one for the ansatz. The Hamiltonian IR consists of a single list of float values containing the number of qubits, the coefficients, and the operators in each Pauli word. The operators are also stored numerically in the list using the binary symplectic vectors \cite{nielsen2010quantum}. The PennyLane quantum chemistry database \cite{arrazola2021differentiable} is used as a baseline to construct the Hamiltonian IR. 
In addition, we developed a set of parsers for each simulator that map the Hamiltonian's IR to the simulator's respective format. 
As for the ansätze IR, we use OpenQASM v2.0 \cite{cross2022openqasm}, which is one of the established IRs for quantum circuits. Then, we parse the IR strings to each simulator, in this way ensuring that the same circuits are running across different simulators; the development of the related code is based on the work from \cite{Gangapuram2024}. Future work includes adapting various other IRs for quantum circuits, including OpenQASM v3.0 \cite{cross2022openqasm} that creates a more coherent environment for representing and handling parametric quantum circuits.

These two IRs ensure a fair comparison of the benchmarks where each simulator uses identical fundamental pieces to enact the identical computational task. 

The software we developed generates all the necessary code to run the simulation. It will soon be made available via the Gitlab platform and as a standalone Python package.


\section{HPC systems}
\label{sec:systems}

The HPC hardware employed in this work is representative of systems users can find in typical computing centers, namely multi-node clusters, shared-memory systems, and GPU-accelerated architectures. 

SuperMUC-NG\footnote{\url{https://doku.lrz.de/supermuc-ng-10745965.html}} (henceforth SNG) is the flagship HPC system of the LRZ and consists of 6336 two-socket Intel\textsuperscript{\textregistered}\  Xeon\textsuperscript{\textregistered}\  Platinum 8174 (code-named Skylake) nodes, each with 48 CPU cores and equipped with 96 GB of main memory.

Eviden\textsuperscript{\textregistered} Qaptiva800\footnote{\url{https://doku.lrz.de/atos-qlm-10745934.html}} is a single-node shared-memory system designed to simulate quantum circuits of up to 38 qubits. It contains 8 Intel Xeon Platinum 8260L (code-named Cascade Lake) CPUs, amounting to a total of 192 cores, with 6 TB of main memory.

The LRZ Big-Data AI Cluster (henceforth BDAI)\footnote{\url{https://doku.lrz.de/lrz-ai-systems-11484278.html}}, mounting NVIDIA\textsuperscript{\textregistered} DGX systems (referred to as DGX), servers specialized in using GPU to accelerate HPC and deep learning applications; a single DGX is equipped with 8 GPUs. 
The DGX systems available for quantum computing simulation are only 
those compatible with the NVIDIA cuQuantum library \cite{the_cuquantum_development_team_2023_10068206}: for our purposes, we used 1 DGX A100, the 3rd generation of DGX, coming with 512 GB memory and 8 Ampere architecture-based A100 GPUs, each with a memory of 80 GB.

\section{Simulation setup}
\label{sec:sim}

\begin{table}[tbp]
    \caption{List of the software simulators used in this study, with the indications of the HPC systems which they have been run on}
    \begin{center}
    \begin{tabular}{l|c|c|c}
        \textbf{Simulator} & \textbf{SNG} & \textbf{Qaptiva800} & \textbf{BDAI} \\
        \hline
        Cirq v1.2.0     & \checkmark & $\times$ & $\times$ \\
        CUDA-Q v0.8.0      & $\times$ & $\times$ & \checkmark \\
        IntelQS v2.1.0  & \checkmark & $\times$ & $\times$ \\
        myQLM v1.10.2   & $\times$ & \checkmark & $\times$ \\
        PennyLane v0.32 & \checkmark & $\times$ & $\times$ \\
        PennyLane Lightning v0.31 & \checkmark & \checkmark & \checkmark\\
        Qiskit v0.45.3  & \checkmark & \checkmark & \checkmark \\
        \hline
    \end{tabular}
    \label{tab:sim-cluster}
    \end{center}
\end{table}

The simulations have been run in different configurations, because different features of the runs and their properties were investigated in different use cases.

For the first use case (\htwo), we used only SNG. The simulations ran in containers as well as bare metal, on a single thread. Each simulation was executed with two different Python interpreters and package sets. One set had interpreters and simulators built with GNU C and C++ compilers; in the other, simulators and interpreters were built with Intel compilers, in an attempt to obtain the best from the underlying architecture.

In the second use case (MaxCut), we used the three HPC systems described in Section~\ref{sec:systems}. 
These tests are focused on a single HPC node, without making use of MPI for node-to-node communication. Depending on the available (or more performant) parallel programming models, either MPI or OpenMP, or a combination of both, has been used. For Qiskit and PennyLane Lightning, support on GPUs is provided by the NVIDIA cuQuantum SDK \cite{the_cuquantum_development_team_2023_10068206}. 
When OpenMP was used, the thread affinity was set through \texttt{OMP\_PLACES=cores} and 
\texttt{OMP\_PROC\_BIND=close}.
Table~\ref{tab:sim-cluster} displays on which cluster each simulator has been executed. 

\begin{table}[htbp]
    \caption{Parallel programming models used for each simulator}
    \begin{center}
    \begin{tabular}{l|c|c|c}
         \textbf{Simulator }& \textbf{OpenMP} & \textbf{MPI} & \textbf{cuQuantum} \\
         \hline 
         Cirq                & $\times$   & $\times$   & $\times$ \\
         CUDA-Q              & $\times$   & $\times$   & \checkmark \\
         IntelQS             & \checkmark & \checkmark & $\times$ \\
         myQLM               & \checkmark & $\times$   & $\times$ \\
         PennyLane           & $\times$   & $\times$   & $\times$ \\
         PennyLane Lightning & \checkmark & $\times$ & \checkmark \\
         Qiskit              & \checkmark & $\times$ & \checkmark \\
         \hline
    \end{tabular}
    \label{tab:sim-parallel-app}
    \end{center}
\end{table}

\section{Simulation software}
\label{sec:software}

The software simulators for this study have been selected based on their community adoption; the list is representative (but far from being complete) of the preferences of typical QC users and not based on an already proven HPC performance. 

In our suite, we included IBM’s Qiskit \cite{Qiskit2023}, Xanadu’s PennyLane \cite{bergholm2018pennylane} (vanilla, with the Lightning Qubit and Lightning GPU plugins), Google’s Cirq \cite{cirq_23}, Eviden’s myQLM\footnote{For simplicity we call "myQLM" both the proprietary Qaptiva simulator software from Eviden which runs on the Qaptiva800 appliance and the standalone software package referenced in \cite{myqlm2024}, which however has not been used in this study.} \cite{myqlm2024}, Intel-QS \cite{ssa16,Guerreschi2020} and NVIDIA’s CUDA-Q \cite{the_cuda_quantum_dev_team}. 

Our benchmarks are focused on using simulators available in the Python ecosystem; the rationale behind this choice is the ease and widespread use of such a language in the scientific community. However, of the above-listed simulators, both IntelQS and CUDA-Q provide an interface to Python, while the simulator is completely written in C++; moreover, PennyLane with the Lightning Qubit plugin uses a backend fully written in C++ (see Table~\ref{tab:sim-cluster} for simulator versions). This difference is worth noting since C++, being a compiled programming language, is generally considered to be faster than an interpreted language like Python; hence, we expect C++-developed simulators to have better performances with respect to fully Python-developed simulators.

While the version of the different software packages is already listed in Table \ref{tab:sim-cluster}, in Table \ref{tab:sim-parallel-app},  the different parallel programming models used in the packages are listed. The only exception is Cirq, which does not natively implement MPI or OpenMP and, therefore, is only used in the \htwo use case, which is run on a single thread. For PennyLane, we use for the parallel tests the Lightning version, which has been developed with focus on HPC performance.

\section{Simulation results}
\label{sec:results}
Our focus is to compare simulators in terms of what is defined in the following as "quantum time," that is, that fraction of the execution time employed in running the quantum circuit: it is a matter of interest, and it takes the majority of the execution time in VQA; the optimization time takes a minor fraction of the whole execution time and thus can be largely ignored. To increase our statistics, we run each of the parameterized circuits 1000 times, each time with different initial values for the parameter set, in this way exploring a large fraction of the parameter space.

\subsection{Use case 1: Molecule Simulation \htwo}
\label{sec:vqe-results}

For this use case, we ran the simulations only on the SNG cluster, both on bare metal and inside a container via the CharlieCloud v0.30 container engine \cite{charliecloud2017}. Here, we provide the results of the profiling delivered by Intel Application Performance Snapshot (Intel APS) v2021.7.1, which is available on SNG, and we compare the execution time on container and on bare metal.

\subsubsection{Profiling results}
\label{sec:profiling-results}
Among the profiling information provided  by Intel APS, we focus on the following quantities:
\begin{itemize}
    \item percentage of \textit{code vectorization},  defined as the fraction of vector instructions issued by the compiler, so capable of being executed in Single Instruction Multiple Data (SIMD) parallel processing;
    \item \textit{memory stalls}, defined as the fraction of slots where the pipeline is stalled due to demand load or store instructions; 
    \item \textit{elapsed time} of the single circuit execution as a proxy for the quantum time because the optimization time has been found to be negligible. 
\end{itemize}
CUDA-Q and  myQLM were not employed in this use case because they primarily target HPC systems different than SNG (the BDAI and Qaptiva800, respectively). 

Regarding the vectorization and memory stalls measures, the percentage of code vectorization is low overall, with a maximum of slightly more than $1.2\%$ for PennyLane with the Lightning Qubit plugin. 

The percentage of memory stalls among accessed memory is around $20\%$ for all simulators. Given the small problem size, the memory footprint is so low 
that all data presumably resides in the cache memory of the CPU. As a consequence, there is little, if any, use of system memory, and cache misses mainly cause the memory stalls reported by Intel APS.

\begin{figure}[tbp]
\centering
\includegraphics[scale=.5]{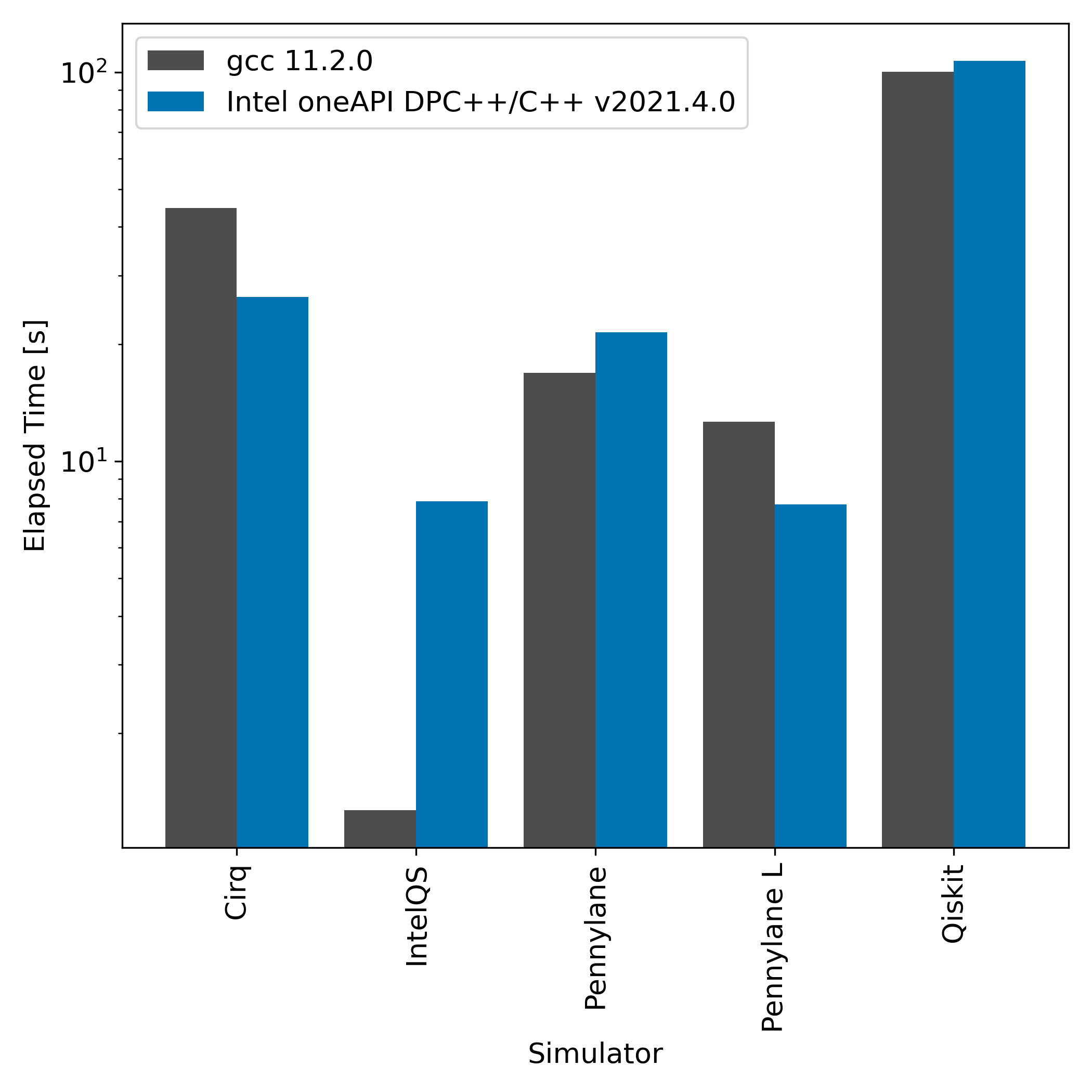}
\caption{Bar plot of the elapsed time to execute the VQE algorithms of the \htwo use case, as reported by the Intel Application Performance Snapshot profiler. The dark gray bars are related to runs executed with Python packages compiled with GNU compiler, while  the blue bars are related to packages built with Intel C++ compiler and run with an Intel-optimized Python interpreter.}
\label{img:elapsed_time_h2}
\end{figure}

Figure~\ref{img:elapsed_time_h2} shows the Intel APS measurements on the elapsed time by reporting the median elapsed time for the execution of the whole VQE simulation, thus including parameters optimization and quantum circuit execution; the median is calculated over 1000 executions. 
Furthermore, the two sets building the bar graph show results from simulators built with GNU compilers (\verb|gcc| v11.2.0, dark grey bars) and simulators built with Intel compilers (Intel oneAPI DPC++/C++ Compiler v 2021.4.0, blue bars); in this last case, we also used the Python interpreter built and provided by Intel which is optimized to perform at best on Intel architectures. With this in mind, the most striking result in Figure~\ref{img:elapsed_time_h2} is the median elapsed time shown by Intel-QS simulations. While staying faster in both scenarios, thus confirming expectations for a C++-based code, Intel-QS shows a smaller elapsed time when using the GNU-built simulator, and the standard Python interpreter is not optimized for Intel architecture.
The simulator exploiting the most from the optimization to the underlying architecture is PennyLane with the Lightning Qubit plugin.

\subsubsection{Container vs Bare Metal}
\label{sec:container-vs-bare-metal}
An additional study we conducted compared the elapsed time when running a simulation on a container to running it on bare metal. The main motivation of this test is to verify the performance of container-run simulations since deploying software on containers is a strategy getting traction in computing centers for its simplicity. The results showed that all the runs on the container were a few percent (< 10\%) faster than the runs on bare metal, independently of the simulator, the compiler, or the interpreter used. This behavior is counter-intuitive (one would expect some kind of overhead by the container machinery) but indeed has already been observed in other studies\cite{bba21}. A possible explanation worth further investigation is related to the thermal design power (TDP) of the CPU, which presumably leads to the application being run at two different clock frequencies on bare metal and in containers.  

\subsection{Use case 2: Optimization Problem - MaxCut}
\label{sec:maxcut-results}
In this use case, the trajectory quantum time $q$ is defined as the median time (averaged over the computed 1000 random realizations of the initial parameters) taken to execute the QAOA circuits needed for the optimizer's convergence. 
This quantity is the central performance metric of this Section. Two different problem sizes have been run, namely graphs with 15 and 20 vertices. The simulations consist of as many qubits as vertices in the graph. For each simulator, we used the parallelization approaches reported in Table~\ref{tab:sim-parallel-app}. 

In what follows, we present the results in terms of \textit{speedup}, defined as the ratio between the computational times running on one thread and the computational time running on $N$ threads. 

\begin{equation}
    \textrm{Speedup} = \frac{t(1)}{t(N)}
\end{equation}

\subsubsection{PennyLane}
\label{sec:pennylane-maxcut-results}
PennyLane, with the so-called Lightning Qubit plugin, shows no scaling with OpenMP as a parallelization scheme, neither on SNG nor on Qaptiva800. The package provides a  software component that would enable parallelization via the adjoint differentiation method \cite{jones2020}, just by setting the number of OpenMP threads before running the simulation. However, such a method cannot be used in our case since it does not support Hamiltonian observables~\cite{pennylaneadjoint}, which are those used in our code.

\subsubsection{myQLM}
\label{sec:myqlm-maxcut-results}
\begin{figure}[tbp]
    \centering
     \includegraphics[scale=0.55]{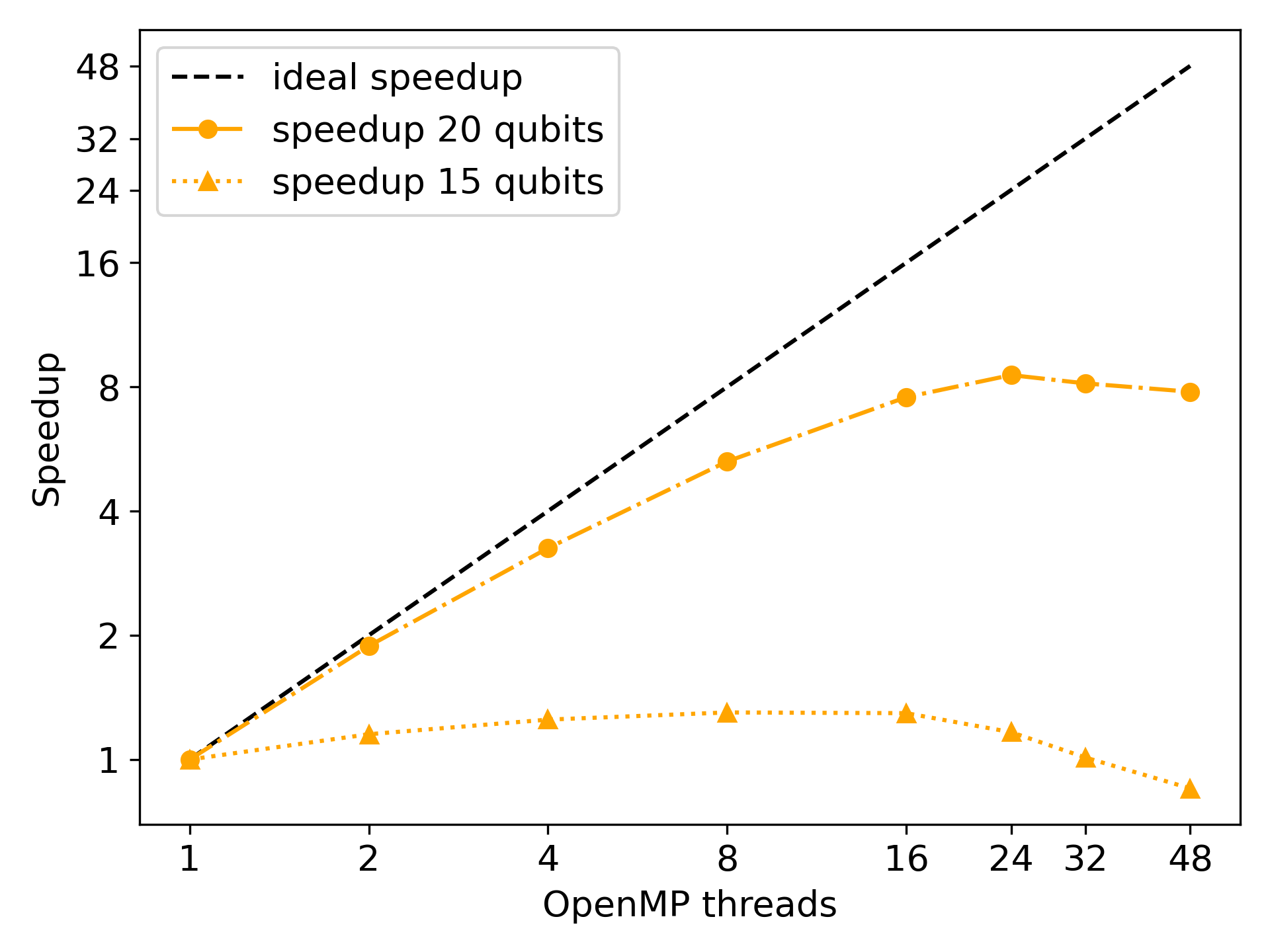}
    \caption{Speedup plot for MaxCut simulations run with myQLM, with 15 and (triangles, dotted line) and 20 qubits (circle, dashed-dotted line) problem sizes. The dashed line shows the ideal speedup. }
    \label{img:myqlm-speedup}
\end{figure}

The median trajectory quantum times for the Eviden simulator are used to produce the speedup plot in  
Figure~\ref{img:myqlm-speedup}. 
Being myQLM proprietary and installed by Eviden on the Qaptiva800 system, the simulations have been run on the Qaptiva800, and the parallelization method used is OpenMP. For the sake of better comparison with SNG runs of other simulators, up to 48 threads are tested.

For the problem on 15 qubits, 
Figure~\ref{img:myqlm-speedup} shows unsatisfactory parallel scaling. The reason for this behavior is most likely the size of the problem, which does not expose enough computation to be parallelized at large numbers of OpenMP threads.

Turning to 20 qubits, the observed speedup follows the ideal speedup line more closely than the 15 qubits case and departs from it only at 24 OpenMP threads. The better scaling of this problem size confirms that, on larger problems, OpenMP can be profitably used for this simulator. 

\subsubsection{Qiskit}
\label{sec:qiskit-maxcut-results}

\begin{figure}[t!bp]
    \centering
    \includegraphics[scale=0.53]{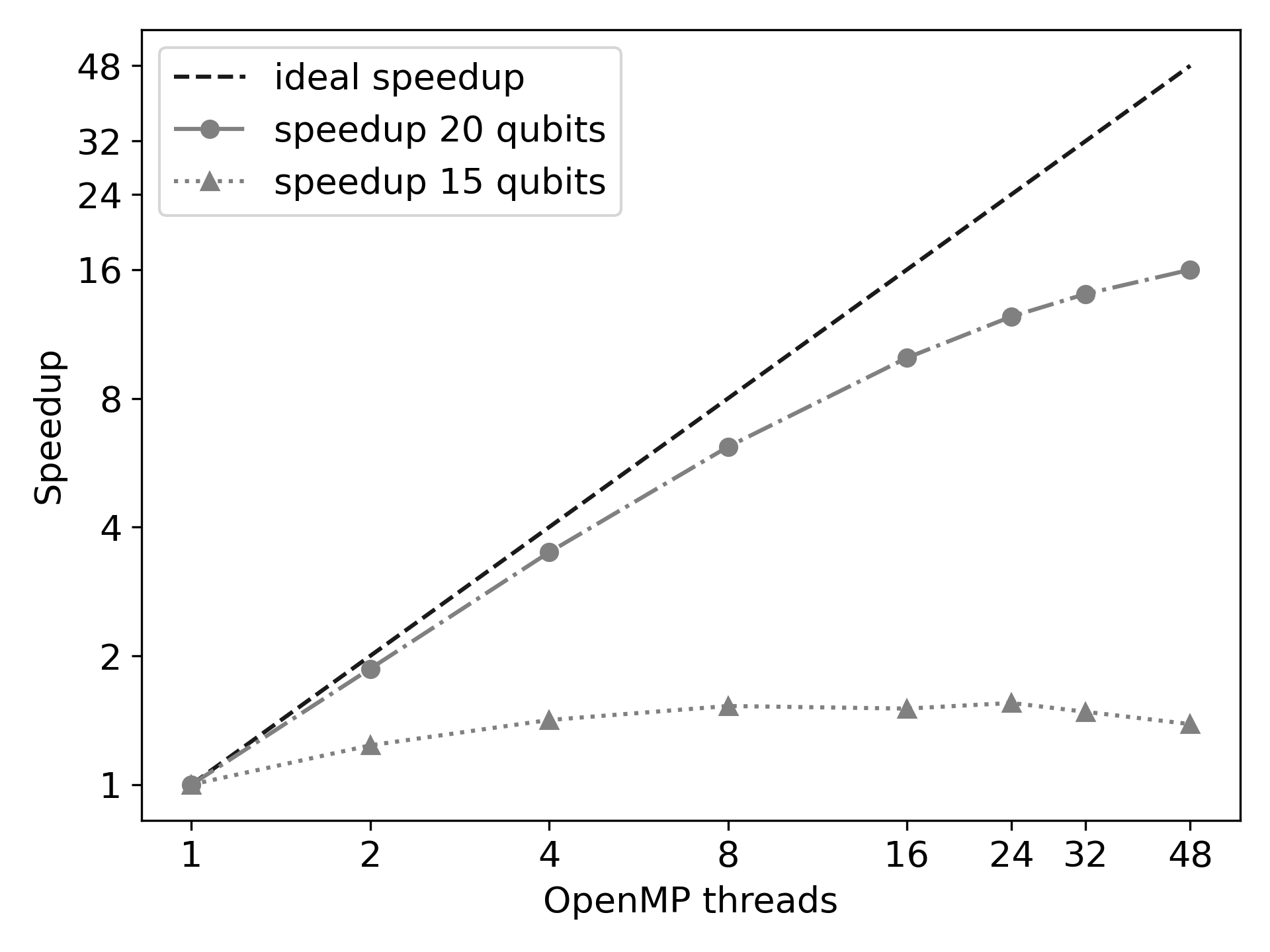}
    \caption{Speedup plot for MaxCut simulations using Qiskit on SNG in both the 15 (triangles, dotted line) and 20 qubits (circle, dashed-dotted line) problem sizes. The dashed line shows the ideal speedup.}
    \label{fig:qiskit-maxcut-speedup}
\end{figure}

Speedup as a function of the number of OpenMP threads for both 15 and 20 qubits on SNG is displayed in Figure~\ref{fig:qiskit-maxcut-speedup}; the dashed line shows the ideal speedup. Similarly to what is described in Section~\ref{sec:myqlm-maxcut-results}, there is very little scaling when solving MaxCut on a 15-node graph. 
Going to the 20-node graph, the speedup is much improved and is better than the one shown by myQLM at a large number of threads. On the other hand, as will be seen in the following, the trajectory quantum time for myQLM is smaller than for Qiskit. 

Concerning the runs on the Qaptiva800 system, Qiskit does not show any speedup; given the behaviour observed in SNG, this was not expected and will be subject of further investigation.

\subsubsection{Intel-QS}
\label{sec:myqlm-maxcut-results}

\begin{figure}[htbp]
    \centering
    \includegraphics[width=0.48\textwidth]{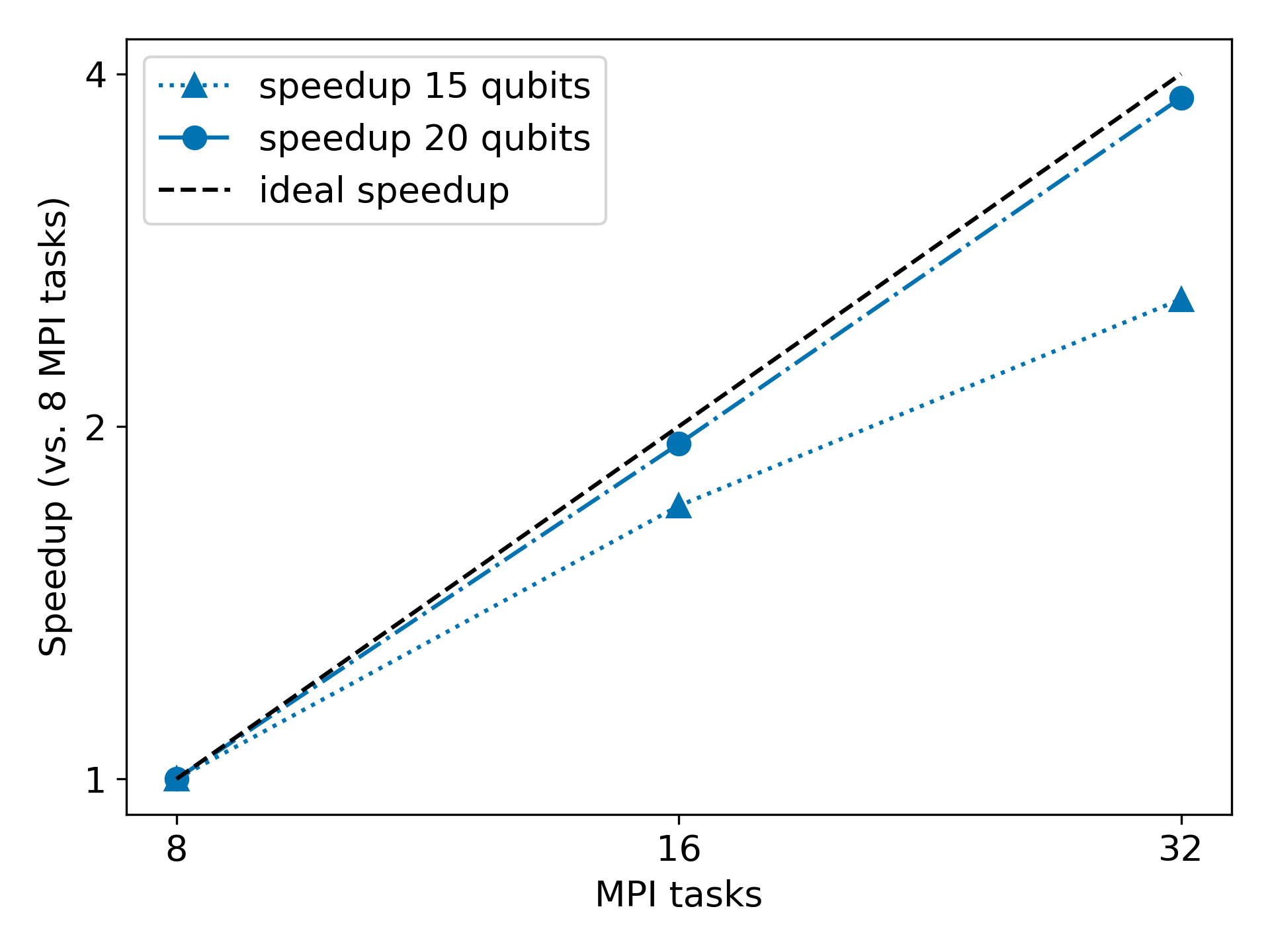}
    \caption{Speedup of simulations for the MaxCut problem using the Intel-QS on 15 and 20 qubits. Simulations were run with a single OpenMP thread and varying numbers of MPI tasks on a single node of the SNG cluster. All speedup values are normalized relative to the performance at 8 MPI tasks.}
    \label{img:intel_qs_mpi}
\end{figure}

\begin{figure}[htbp]
    \centering
    \includegraphics[width=0.48\textwidth]{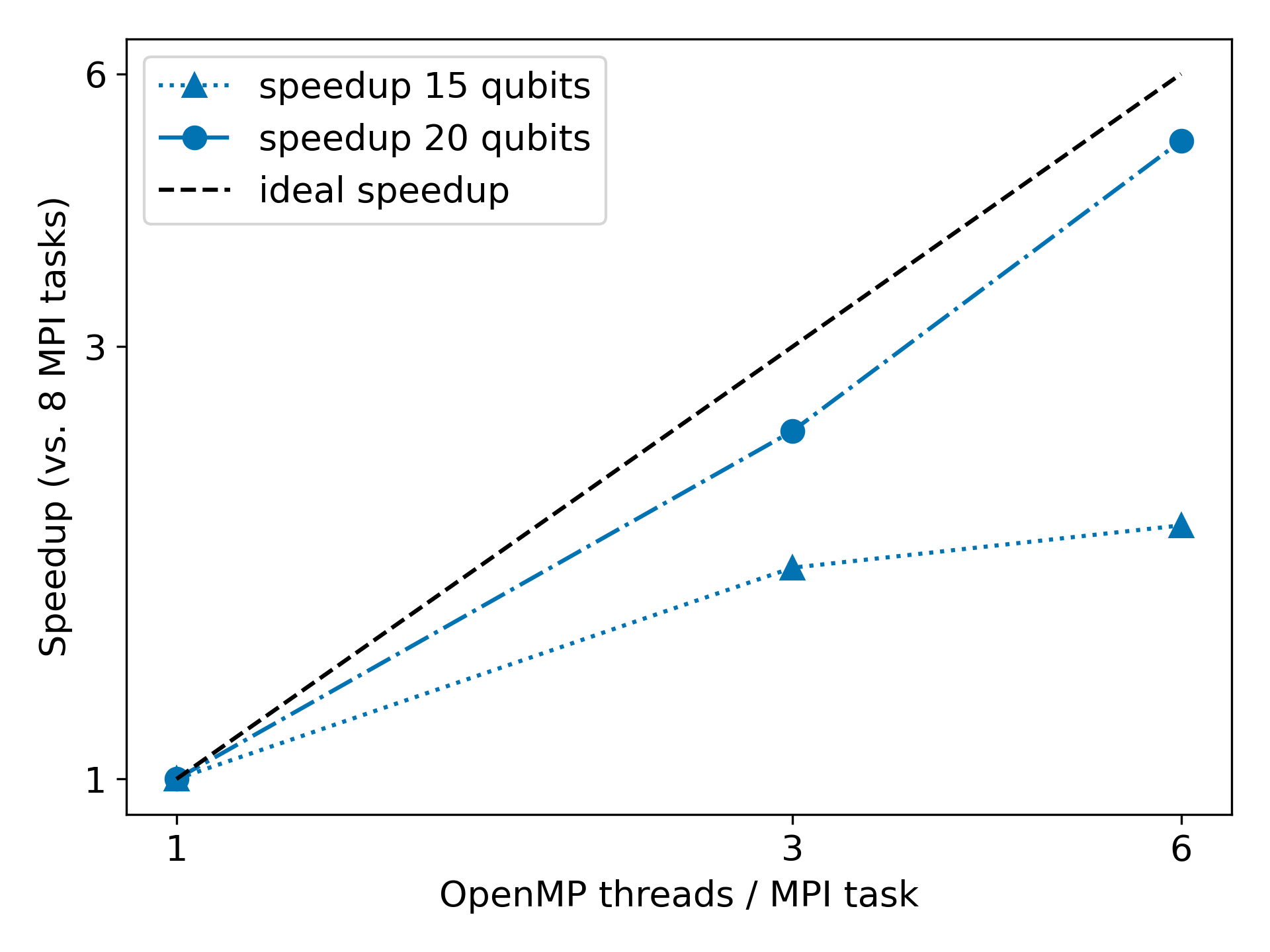}
    \caption{Speedup of simulations for the MaxCut problem using the Intel-QS on 15 and 20 qubits. Simulations were run with 8 MPI tasks and varying numbers of OpenMP threads on a single node of the SNG cluster. All speedup values are normalized relative to the performance at 8 MPI tasks.}
    \label{img:intel_qs_omp}
\end{figure}

Since Intel-QS supports both distributed and shared-memory parallelism, this Section evaluates the efficiency of parallel computations using pure MPI and hybrid MPI/OpenMP configurations on SNG.
Figure \ref{img:intel_qs_mpi} presents the speedup for Intel-QS simulations with 15 and 20 qubits, based on the median of 1000 execution times for each point. The speedup is shown as a function of the number of MPI tasks without using OpenMP.
An ideal scaling line, normalized to the 8 MPI task configuration, is also included for reference. 
The problem on 15 qubits shows the same limitations at large number of parallel workers seen with the other simulators. Conversely, the scaling looks close to ideal for the problem on 20 qubits.
This further suggests that as the problem size increases, the computational workload better amortizes the communication cost, improving parallel efficiency. With respect to the codes seen before, Intel-QS has a limitation in its parallel strategy: the number of MPI tasks can only have values that are powers of 2.

To further investigate parallel performance, Figure \ref{img:intel_qs_omp} explores an alternative configuration in which the number of MPI tasks is fixed at 8, and the number of OpenMP threads per task is increased from 1 to 6. This setup allows us to assess the effectiveness of hybrid parallelism on a single node of the SNG cluster, utilizing all available CPU cores at the 6-thread configuration.

Like the trends observed in Figure \ref{img:intel_qs_mpi}, speedup increases as more computational resources are employed. However, the efficiency of scaling with OpenMP threads is noticeably lower compared to scaling with MPI tasks. This inefficiency is especially evident in the 15-qubit case, where the execution time with 6 threads diverges significantly from the ideal speedup curve. This suggests that thread management and synchronization overheads for smaller problem sizes outweigh the benefits of additional threads. In contrast, for the 20-qubit case, scaling behavior improves, and the performance remains closer to ideal as the increased computational workload better utilizes the available resources.

Comparing Figures \ref{img:intel_qs_mpi} and \ref{img:intel_qs_omp}, it can be concluded that, for the QAOA circuit simulated in this study, hybrid MPI/OpenMP parallelism does not offer clear advantages over a pure MPI approach. While increasing the number of OpenMP threads per task generally reduces execution time, it does not scale as efficiently, particularly for smaller problem sizes. However, the combination of MPI and OpenMP fully utilizes the available 48 cores of an SNG node and outperforms the MPI-only run with 32 tasks.

To sum up the results of the CPU-only runs, the quantum time for simulations of MaxCut on 20 qubits on 48 OpenMP threads (equivalent to a full SNG node), or a suitable combination of MPI and OpenMP (for Intel-QS) are reported in Table~\ref{tab:qtime-48}. We observe that the timings of Pennylane are much larger than the others, as explained by the missing parallelization in this code for the formulation chosen for our problem. On SNG, Intel-QS slightly outperforms Qiskit. As the former code relies on an HPC software stack, it was not possible to install and test it on the Qaptiva800; on that system, Qiskit runs faster than on SNG (because of the newer generation of Skylake cores) but considerably slower than myQLM, which is the fastest on that system. One should add that, for completeness, on the Qaptiva800, myQLM has been forced to run with a user-defined number of OpenMP threads rather than resorting to its heuristics to ease the comparison with the other codes.

\begin{table}[htbp]
    \caption{Quantum time for the MaxCut simulations, 20 qubits, on CPU, in seconds.}
    \begin{center}
        
    \begin{tabular}{l|c|c}
         \textbf{Simulator} & \textbf{SNG} & \textbf{Qaptiva800} \\
         \hline
         PennyLane Lightning & $0.9275$ & $0.6290$ \\
         myQLM & -- & $0.0702$ \\
         Qiskit & $0.1136$ & $0.1099$ \\
         IntelQS & $0.1004$ & -- \\
         \hline
    \end{tabular}
    \label{tab:qtime-48}
    \end{center}
\end{table}

\subsubsection{CUDA-Q}
\label{sec:cudaq-maxcut-results}

\begin{figure}
\includegraphics[scale=0.48]{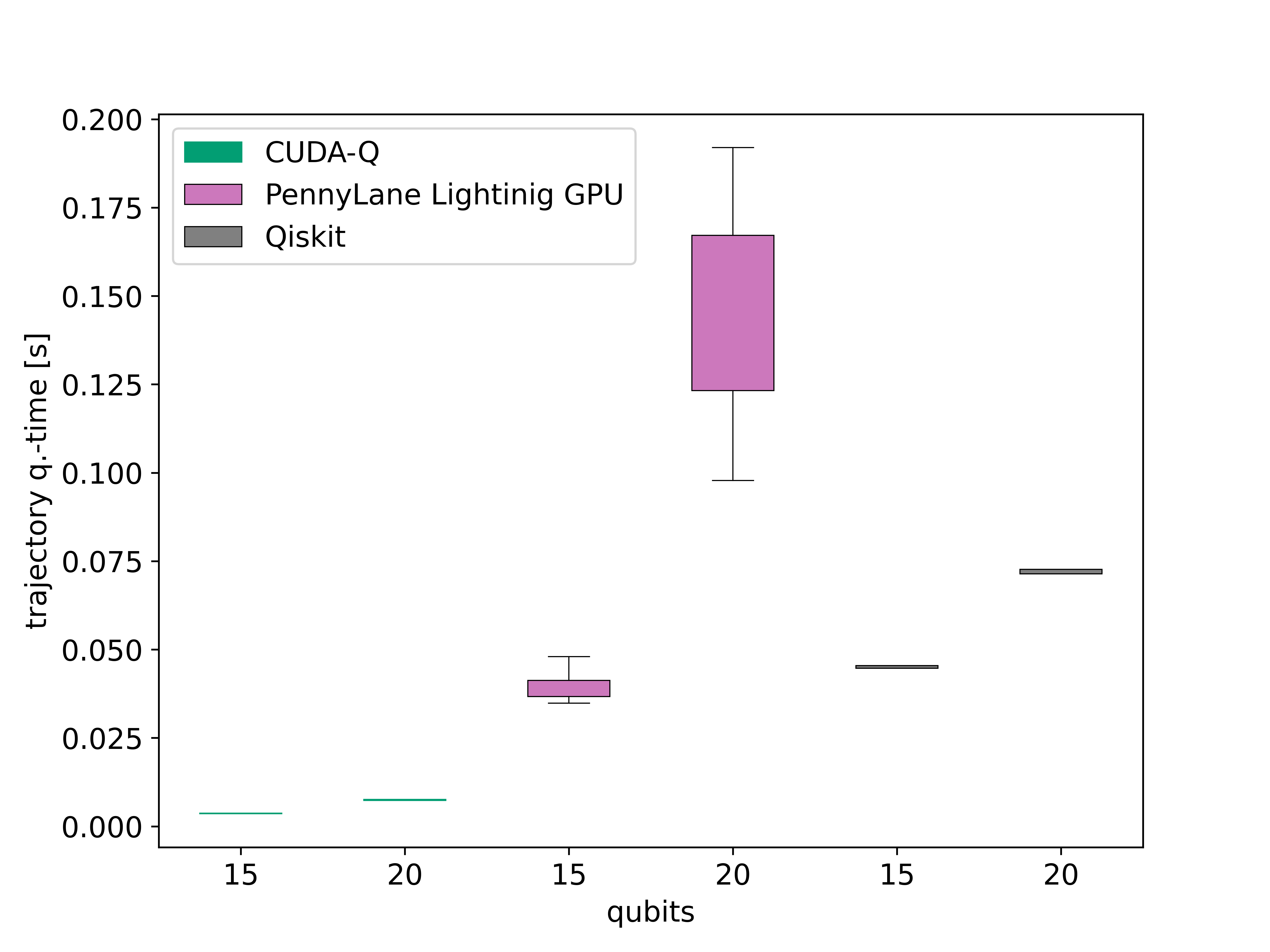}\textbf{}
\caption{Box plots of the $q$ time for different simulators and problem sizes on the BDAI system. We removed the whiskers from the Qiskit and CUDA-Q boxplots to improve the appearance. }
\label{img:AI_maxcut_qtimes}
\end{figure}

Let $Q_{n,\text{sim}}$ be the median trajectory quantum time for the $n$ qubit MaxCut instance for a specific quantum simulator.
Let 

$S_{\text{sim}}=Q_{20,\text{sim}}/Q_{15,\text{sim}}$ 
be the scaling factor from the 15- up to the 20-qubit size problem for a given simulator.  
Consider that, going from 15 to 20 qubits, the system size is up to 32 times larger in the worst case. Keeping this in mind, the results of the trajectory quantum time on the BDAI system (Figure \ref{img:AI_maxcut_qtimes}) indicate that the observed scaling factors $S_{\text{sim}}$  scale with the problem size but well below the worst case by a large margin. This indicates that cuQuantum, composed of libStateVec and libTensorNet, developed in CUDA~\cite{Kirk2007}, uses the parallelization potential of the problem very well. The hardware vendor designs these softwares to work at best on its GPUs; the fact they show the best scaling is to be expected.
Considering this, \mbox{CUDA-Q} demonstrates the overall best performance. The lower performance of PennyLane Lightning GPU and Qiskit relative to \mbox{CUDA-Q} indicates that the software  interfaces to the underlying simulation backend cuQuanutm introduce some overhead. A potential analysis of the larger scattering of the trajectory quantum times for PennyLane Lightning GPU is deferred to future work. The GPU-accelerated simulators scale with a factor of $S_{\text{CUDA-Q}}\approx 1.7$, $S_{\text{PennyLane Lightning GPU}}\approx 3.3$ and $S_{\text{Qiskit}}\approx 1.6$ from the 15 to the 20 qubit MaxCut instance.

A comparison of the data in Figure \ref{img:AI_maxcut_qtimes} with the performance numbers in Table \ref{tab:qtime-48} is not entirely fair without considering the peak performance of all involved computing systems. However, while results for Qiskit and PennyLane on BDAI are in line, within a factor of a few, with the findings on the CPU systems, the quantum time of CUDA-Q stands out. It shows that they are a valuable tool to perform quantum circuit simulations.

\subsection{Consistency check and use case 3: TSP}
\label{consistency-check}
The reported results are validated through consistency checks. The optimization procedure results are utilized to verify the equivalence of the different implementations and the Hamiltonian and ansatz parsers. They also serve as a sanity check of the different simulators' methods of simulating the circuit and then computing expectation values. This is done for the MaxCut problem, which was solved using the QAOA algorithm and use case 3, TSP, with VQE.

\begin{figure}[h]
    \centering
    \includegraphics[width=0.45\textwidth]{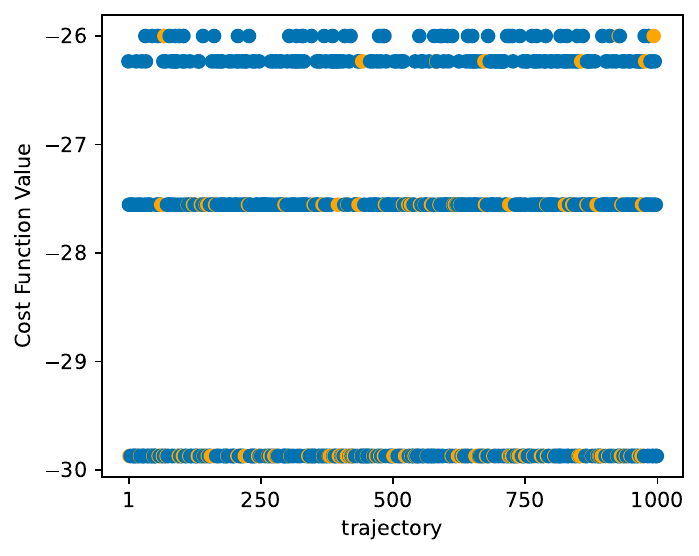}
    \caption{Cost function value for each trajectory for execution with PennyLane for MaxCut with 15 qubits. Blue dots represent trajectories for which the optimizer reports a successful convergence, while orange dots report the opposite scenario.}
    \label{fig:energies_pennylane_15q}
\end{figure}

The QAOA algorithm does not find the optimal solution in all cases, but the results for the different trajectories with varying starting points accumulate on discrete levels, indicating local minima in the landscape of the cost function. This is shown exemplarily for PennyLane Lightning Qubit in Figure~\ref{fig:energies_pennylane_15q}. The classical optimizer finds these levels with high precision as the variance within these levels is in the order of $10^{-13}$ for 15 qubits and $10^{-3}$ for 20 qubits. However, there is no unique point in the loss landscape corresponding to a particular cost function value, and the parameters of the ansatzes show accumulation at several points. For some trajectories, the optimization routine does not report successful convergence despite resulting in the same final cost function values as for successful trajectories. The ratio of the trajectories with reported successful convergence depends on the simulator used and is listed in table \ref{tab:convRatioMaxcut}.

\begin{table}[htbp]
    \caption{Ratio of runs with successful convergence (CR) as reported by the optimizer for the different simulators for the MaxCut problem with 15 and 20 qubits.}
    \begin{center}
        
    \begin{tabular}{l|c|c}
         \textbf{Simulator} & \textbf{CR 15 qubits}  & \textbf{CR 20 qubits} \\
         \hline
         PennyLane Lightning & $0.841 $ & $0.427$ \\
         myQLM & $1.0 $ & $0.95$ \\
         Qiskit & $0.846$ & $0.417$ \\
         IntelQS & $0.98$ & $0.933$ \\
         \mbox{CUDA-Q} & $0.851$ &  $0.41$ \\
         \hline
    \end{tabular}
    \label{tab:convRatioMaxcut}
    \end{center}
\end{table}

\begin{figure}
    \centering
    \includegraphics[width=0.45\textwidth]{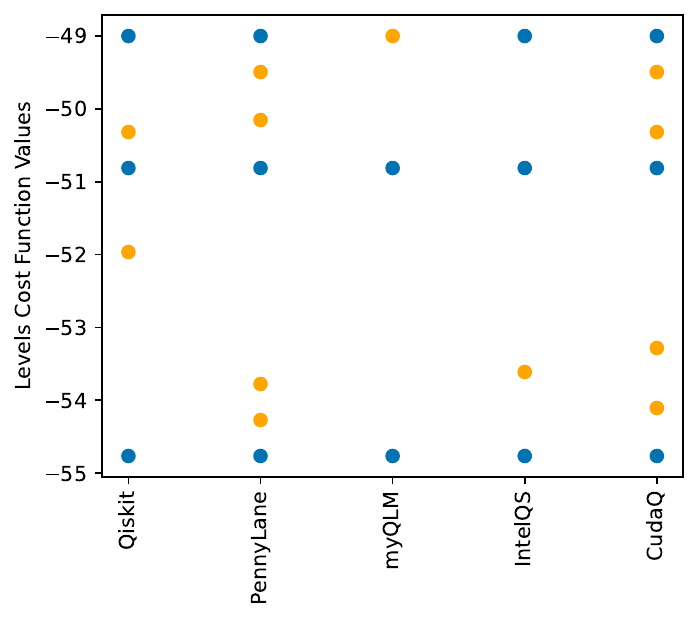}
    \caption{Centers of the levels of the cost function values for the different simulators for the MaxCut problem on 20 qubits. Yellow markers indicate levels with less than 30 points.}
    \label{fig:energyLevelsMaxcut20}
\end{figure}

This behavior is reproduced for the different setups reported before. 
Figure \ref{fig:energyLevelsMaxcut20} shows the values of the discrete levels for the 20-qubit problem instance for the different simulators. Orange markers represent groups with less than 30 points, i.e. sparsely populated levels. For the 20 qubit problem, there are some events that are additional to the main levels. Together with the lower convergence ratios, this is a symptom of the more complicated cost function landscape. The optimization results can be susceptible to several factors, including the use of single or double precision. In none of the iterations, the global minimum at $-34$ and $-68$ is found for the 15 and 20 qubit-problem, respectively.  
Figure \ref{fig:energyLevelsMaxcut20} displays that the main cost function levels align perfectly for the different simulators in the example of the problem instance with 20 qubits. 

The state vector simulations for VQE on 4 node instances, i.e. 9 qubits, are performed for the TSP use case. Similarly to Figure~\ref{fig:energies_pennylane_15q}, 
the cost function values accumulate on discrete levels. 

Most simulators agree with each other, with energy differences at the level of numerical round-off ($\sim 10^{-12}$ absolute energy difference). \mbox{CUDA-Q} does not converge to the same values, but with a negligible discrepancy at the level of $10^{-5}$ off the value. An investigation of the source of this issue and the whole optimization process shows that $32.1\%$ of runs converge to the optimal compared to $34.7\%$ for all other simulators. The small difference in the average cost function value stems from the expectation value computation in the \mbox{CUDA-Q} package, as the actual state vectors for the optimization results are identical.

\section{Discussion and conclusions}
\label{discussion}

Several approaches are possible when comparing the performance of different simulators. One of them is to describe the problem and its initial conditions and let the different codes solve it with their own distinct methods (for a textbook example of it in astrophysics, see \cite{fwb99}). What we designed in this study is somewhat orthogonal and, using a Hamiltonian and ansatz parser, is aimed to pass precisely the same circuits to all simulators. The verification run in Section \ref{consistency-check} proved this approach successful, thus providing the user community with a valuable tool to port problem instances in optimization problems between different simulators. 

Besides that, the performance results have shown that the use cases presented here all share challenges, from the HPC performance viewpoint. They range from the widespread usage of Python for such tools to unsatisfactory vectorization (see Section \ref{sec:vqe-results}) for the \htwo use case. MaxCut is a simulation problem very unbalanced towards long simulation times (because of the need for running many instances of the randomly chosen initial parameters), while exposing relatively small computations to parallelization. The 20-qubit instance of the problem presents a promising scaling on one node, however this is already at the limit of what is possible to simulate usefully in terms of time to solution, and has a memory footprint of its quantum state of only 16 MB, much smaller than the RAM memory available on the node. The problem has already been highlighted \cite{Guerreschi2020} and can be partially addressed by bundling more trajectories in so-called job arrays. We stress that these conclusions can be drawn from HPC simulations even at the small scale of one HPC node, like the ones performed in this study. 

We have focused on codes relevant to their wide adoption in the user community. In the literature, one can find research examples of applications aiming at a higher utilization of HPC resources \cite{xbs23}, but they were not in focus of this study. One noteworthy remark is that simulations on GPUs appear to provide a balanced mix of usability and performance also on community codes. GPU-accelerated HPC machines emerge therefore also in this area as a very suitable tool for simulations.

Future work will focus on techniques for further upscaling the runs studied here and on understanding how simulations on leadership-scale HPC systems can support the progress of algorithms and hardware in quantum computing.

\section*{Acknowledgments}
This work is partially funded by the Bavarian State Ministry of Science and the Arts as part of the Munich Quantum Valley and by the Bavarian Ministry of Economic Affairs, Regional Development and Energy with funds from the Hightech Agenda Bayern.

\bibliographystyle{IEEEtran}
\bibliography{benchcomp}

\end{document}